\documentclass[preprint,12pt]{elsarticle}

\usepackage{graphicx}
\usepackage{epsfig}

\usepackage{amssymb}
\usepackage{amsthm}
\usepackage{amsmath}

\begin{document}

\begin{frontmatter}




\title{The numerical operator method to the real time dynamics of currents through the nanostructures with different topologies}


\author{Pei Wang}
\ead{wangpei@zjut.edu.cn}
\address{Institute of applied physics, Zhejiang University of Technology, Hangzhou, P. R. China}
\author{Xuean  Zhao*}
\ead{zhaoxa@zju.edu.cn}
\address{Zhejiang Institute of Modern Physics, Department of Physics, Zhejiang University, Hangzhou, P. R. China}
\cortext[cor]{Corresponding author: Phone: 0086-571-87953690}
\author{Ling Tang}
\address{Institute of applied physics, Zhejiang University of Technology, Hangzhou, P. R. China}
\begin{abstract}

We present the numerical operator method designed for the real time dynamics of currents through nanostructures beyond the linear response regime. We apply this method to the transient and stationary currents through nanostructures with different topologies, e.g., the flakes of square and honeycomb lattices. We find a quasi-stationary stage with a life proportional to the flake size in the transient currents through the square flakes, but this quasi-stationary stage is destroyed in the presence of disorder. However, there is no quasi-stationary stage in the transient currents through the honeycomb flakes, showing that the transient current depends strongly upon the topologies of the nanostructures. We also study the stationary current by taking the limit of the current at long times. We find that the stationary current through a square flake increases smoothly as the voltage bias increasing. In contrast, we find a threshold voltage in the current-voltage curve through a honeycomb flake, indicating a gap at the Fermi energy of a honeycomb flake.
\end{abstract}


\end{frontmatter}


\section{Introduction}

The coherent transport through nanostructures beyond the linear and static response regime has generated a lot of interest in the past decade. From the experimental point of view, this is due to the development of nanotechnology that permits to fabricate small devices and manipulate the current at a large voltage bias under high frequency. And from a theoretical point of view, this is because transport properties beyond the linear and static response regime explore a quantum system far from equilibrium, which has been poorly understood due to the lack of a variational principle~\cite{glansdorff}.

The usual way of studying the coherent transport beyond the linear and static response regime is by the non-equilibrium Green's function method. Using this method Jauho et al.~\cite{jauho94} solved the transport problems in a quantum dot with two reservoirs for DC and harmonic AC electric voltages. But it is a formidable task to deal with the strong interactions and transient dynamics. The Keldysh formalism is usually expressed in the frequency representation and is then difficult to deal with the real time dynamics of the systems~\cite{rammer}. However, the real-time dynamics of the systems is important for understanding the fast input and output of the solid state qubits and the electron motion in nanostructure devices, which are promising candidates for implementing a scalable quantum computer~\cite{elzerman04,hanson05}. Therefore, there have developed several numerical techniques to study the current dynamics~\cite{muehlbacher08,anders05,meisner09,schoeller09,kennes12a} beyond the Keldysh formalism. Most of these techniques concentrate on the single impurity Anderson model that describes a single level with strong electron-electron interactions~\cite{anders05,meisner09,meir93,werner09}.
Much less is known about the current dynamics as the nanostructure has internal degrees of freedom. Even in the absence of electron-electron interactions, solving the transient time-dependent current through a many-body nanostructure in real space is non-trivial~\cite{kohler05}. Especially at the long time scale the problems become much more difficult. While, the long-time current-carrying state is necessary for understanding the evolution of current from transient state to stationary state, since the relaxation time of the current will increase as the size of a nanostructure increases.

In real electronic devices, such as field-effect transistors, there are two semi-infinite leads connecting the device to the voltage or current contacts. In practical computation of current for longer time, more sites in the leads must be involved. The current cannot be obtained by exact diagonalization of the single-particle eigenmode for a long time evolution. Recently, the real time dynamics of currents through the one-dimensional Hubbard model, which is a model of quantum wires or carbon nanotubes, has been studied by the adaptive time-dependent density matrix renormalization group (tDMRG) method~\cite{meisner10}. However, tDMRG can only be used in one dimension and suffers from the finite-size problem. It is obliged to improve a new numerical method for finite two dimensional structures. In this paper, we will introduce the numerical operator method, which is first developed in Ref~\cite{wang13}. In this method, the current is calculated by solving the Heisenberg equation iteratively. By introducing a truncation scheme, the long-time current state can be obtained. We use this method to study the current dynamics through two kinds of two-dimensional nanostructures, in which the atoms form a square lattice and honeycomb lattice (graphene)~\cite{peres10,pal11,ponomarenko08}, respectively, as shown in Fig.~\ref{fig:model}. To compare the current dynamics in the two different topological structures, we calculate the transient current and stationary current in the perfect and disorder configurations, respectively. We find a quasi-stationary current through a square lattice flake, while it does not exist in a honeycomb lattice flake.  In the presence of disorder, the quasi-stationary current through the square flake is destroyed. We observe a threshold voltage in the I-V curve of a honeycomb flake. The detailed discussions are given in the following sections.

The contents of the paper are arranged as follows. In Sec.~2, we introduce the two different structures and the theoretical model. The numerical operator method is introduced in Sec.~3. In Sec.~4 and~5, we discuss the transient currents without and with disorder respectively. In Sec.~6, we discuss the stationary currents. Sect.~7 is a short summary.

\section{The model}

We consider a model containing two semi-infinite leads and a scattering region (the nanostructure), which is a flake of two-dimensional lattice. The total Hamiltonian is $\hat H = \hat H_C + \hat H_L + \hat H_R + \hat H_V$, where
\begin{equation}
 \hat H_L = -g_l \sum_{j=-\infty}^{-2} (\hat c^\dag_{j} \hat c_{j+1} + h.c.),
\end{equation}
and
\begin{equation}
 \hat H_R = -g_l \sum_{j=N}^{\infty} (\hat c^\dag_{j} \hat c_{j+1} + h.c.),
\end{equation}
describing the left and right leads respectively, $\hat H_C$ the flake and $\hat H_V$ the coupling between the leads and the flake. The $\hat c^\dag_j$ denotes the fermionic creation operator at site $j$, and $N$ the number of sites in the flake.

\begin{figure*}
\begin{center}
\includegraphics[width=0.7\textwidth]{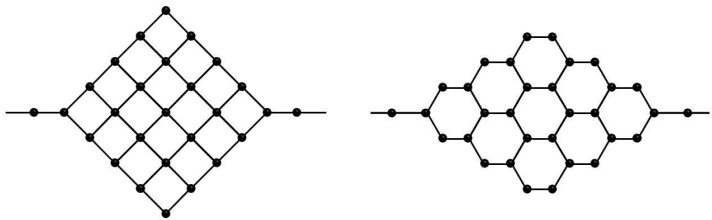}
\caption{\label{fig:model} The schematic diagram of the square and honeycomb flakes. The left figure shows a square flake of size 5x5, and the right one is a honeycomb flake of size 4x4.}
\end{center}
\end{figure*}

Two kinds of flakes are studied in this paper: the square flake and the honeycomb flake (their structures are shown in Fig.~\ref{fig:model}). The square flake is a square-shaped piece cut off from a square lattice, and its two opposite vertices are coupled to the left and right leads respectively. The size of a square flake is given by the number of its edge sites. We use ``$n\times n$" to denote a square flake whose each edge contains $n$ sites. The honeycomb flake is a diamond-shaped piece cut off from a honeycomb lattice with all the four edges zigzag edges. A honeycomb flake of size ``$n\times n$" denotes a flake whose each edge comes from the edges of $(n-1)$ hexagons. The total number of its sites is then given by $N=(2\times n^2-2)$.

The Hamiltonian of the flake is expressed as
\begin{equation}
 \hat H_{C} = -g_d \sum_{\langle i,j\rangle} (\hat c^\dag_i \hat c_j + h.c.) + \sum_j V_g(j) \hat c^\dag_j \hat c_j,
\end{equation}
where $g_d$ denotes the coupling between two neighbor sites of the flake and $V_g(j)=V_g + {\bf \eta} \epsilon_j$ the onsite potentials. The onsite potential contains an overall shift $V_g$ (called the gate voltage) and a disordered term ${\bf \eta} \epsilon_j$ where $\epsilon_j$ is a uniformly distributed random number in the interval $[-\frac{1}{2},\frac{1}{2}]$ and ${\bf \eta}$ the strength of disorder.

The coupling between the flake and the two leads is time-dependent, described by
\begin{equation}
 \hat H_V = g_c(t) \left( \hat c^\dag_{-1} \hat c_0 + \hat c^\dag_{N-1} \hat c_{N} + h.c. \right),
\end{equation}
where the site $0$ and $N-1$ denote the left and right vertices of the flake respectively, and $g_c(t)= g_c \theta(t)$ with $g_c$ the coupling strength and $\theta(t)$ the Heaviside step function. We take the wide band limit by setting $g_l=10 g_c$ throughout the paper. And we set $g_c=g_d$ so that the level spacing of the flake is correspondingly small.

\section{The numerical operator method}

We calculate the time-dependent current through the flake after the coupling between leads and the flake is switched on at time $t=0$. The current is the average of the left and right currents, and is expressed as
\begin{equation}
 I(t) = -g_c \left( \textbf{Im} \langle \hat c^\dag_{-1}(t) \hat c_0(t) \rangle + \textbf{Im} \langle \hat c^\dag_{N-1}(t) \hat c_N (t) \rangle \right).
\end{equation}
The averaged current will be the same as the left or right current in the steady limit, and shows similar features for the transient regime.

To obtain the current, we calculate four field operators $\hat c^\dag_{-1}(t), \hat c_0(t), \hat c^\dag_{N-1}(t)$ and $\hat c_N(t)$ and then their expectation values with respect to the initial state. The field operators satisfy the Heisenberg equation $d \hat c^\dag_j(t)/dt =i [\hat H, \hat c^\dag_j(t)]$, which is solved by the numerical operator method, first developed in Ref~\cite{wang13}. The procedure is summarized as follows.

We suppose that the solution of the Heisenberg equation at the time $t$ can be expressed as
\begin{equation}\label{suppcj}
 \hat c^\dag_j (t) = \sum_k W_{jk}(t) \hat c^\dag_k,
\end{equation}
where $W_{jk}(t)$ is the propagator at the time $t$. This expression is valid since the Hamiltonian is quadratic. Then at the time $t+\Delta t$, we have
\begin{equation}
\begin{split}
\sum_k W_{jk}(t+\Delta t) \hat c^\dag_k = e^{i\hat H \Delta t} \hat c^\dag_j (t) e^{-i\hat H \Delta t}.
\end{split}
\end{equation}
By setting $\Delta t$ very small, the right hand side can be expressed as a power series of $\Delta t$. Keeping to the second order, we have
\begin{equation}\label{seconddeltat}
\begin{split}
\sum_k W_{jk}(t+\Delta t) \hat c^\dag_k = \hat c^\dag_j (t) + i\Delta t[\hat H, \hat c^\dag_j (t)] -\frac{\Delta t^2}{2} [\hat H,[\hat H,\hat c^\dag_j(t)]] .
\end{split}
\end{equation}
Substituting Eq.~\ref{suppcj} into Eq.~\ref{seconddeltat}, and supposing that $[\hat H, \hat c^\dag_k]= \sum_l G_{kl} \hat c^\dag_l $, we obtain
\begin{equation}
\begin{split}
 W_{jk}(t+\Delta t)= W_{jk}(t) +i\Delta t \sum_l W_{jl}(t) G_{lk} -\frac{\Delta t^2}{2} \sum_{l,m} W_{jl}(t) G_{lm} G_{mk}.
\end{split}
\end{equation}
The propagators at the time $t+\Delta t$ are linear functions of those at the time $t$. Then the propagators at arbitrary time can be worked out by an iterative algorithm starting from $t=0$ and moving forward by $\Delta t$ at each step. The initial values of the propagators are $W_{jk}(0)= \delta_{j,k}$.

At the time $t$, we only need to store the present non-zero propagators. At the initial time, a single propagator $W_{j,j}(0)=1$ is stored. But the number of non-zero propagators increases in course of time. To obtain $W_{jk}$ at very large times, we carry out a truncation scheme at each step and keep only $M$ propagators with the largest magnitudes. This truncation scheme extraordinarily increases the largest time that we can reach and at the same time keeps the result in high precision. Errors in our algorithm come from the finite $\Delta t$ and $M$. The error caused by a finite $\Delta t$ is in order $O(\Delta t^3)$ and can be made arbitrarily small by taking $\Delta t \to 0$. The error by a finite $M$ can be reduced by increasing $M$. In practice, setting $M$ to tens of thousands has been able to make the error negligible in the range of time that we can reach.

\section{The transient currents through the square and honeycomb flakes as $\eta =0$}

\begin{figure}
\begin{center}
\includegraphics[width=0.7\textwidth]{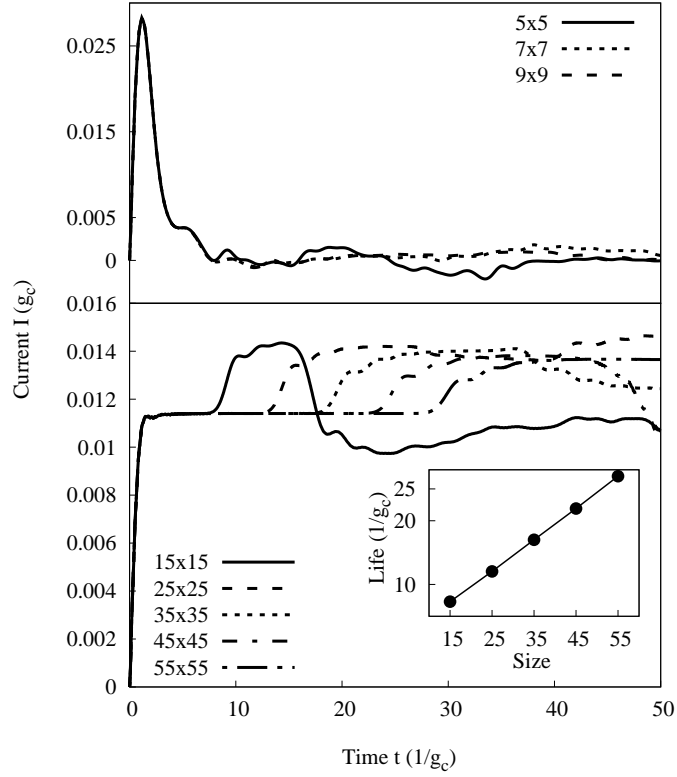}
\caption{[Bottom panel] The transient currents through the square flakes of different sizes denoted by $``n\times n"$ (the meaning of $n$ is introduced in Sec.~2). We find the state with a quasi-stationary current, whose life increases linearly with the flake size. The inset figure shows the ending time of the quasi-stationary state as a function of the flake size. The voltage bias is set to $V=0.5g_c$, and the gate voltage to zero. [Top panel] The transient currents through honeycomb flakes of different sizes. The voltage bias is set to $V=1.4g_c$.\label{fig:squaretransient}}
\end{center}
\end{figure}
In this section, we discuss the transient currents through the square and honeycomb flakes in the absence of disorder. Let us consider the initial condition that at the time $t=0$ the leads and the flake are decoupled to each other, and there is no electrons in the flake, which is realized by applying a sufficiently negative gate voltage in experiments. The leads are in thermal equilibrium with the chemical potentials $\mu_L=V/2$ and $\mu_R=-V/2$ respectively, where $V$ denotes the voltage bias. After the coupling is switched on at the time $t=0$, electrons pour into the flake from two leads. While, the currents from the left and right leads are not the same, resulting in a non-zero net current through the flake.

The bottom panel of Fig.~\ref{fig:squaretransient} shows the transient currents through square flakes of different sizes. We see that the current increases abruptly at the beginning ($t=0\sim7$). At this stage, the current is independent of the size of the flake. Currents are the same for different flakes. This result indicates that electrons injected from the leads just pass through the flakes directly over channels connected the left and the right leads. They do not reach the boundary of the flake. The influence of the boundary does not take effect.  Also for all the square flakes, there are quasi-stationary periods that the transient currents keep the same values, as shown in the bottom panel of Fig.~\ref{fig:squaretransient}. The larger size, the longer quasi-stationary state. The transport through a square flake is a ballistic process, in which all the channels are open. There is no reflected waves from the boundary to interfere the channel electrons. This explains an approximately linear increase of the quasi-stationary life time. When exceeding the quasi-stationary stage, the currents begin to increase rapidly. This is due to the reflected currents from the boundaries.

For a honeycomb flake, there also exists  a time range in which the transient current is independent of the flake size.  This is due to that electron waves have not reached the flake boundary (see the top panel of Fig.~\ref{fig:squaretransient}), and then the boundary effects do not play a role. As time going on, the electrons arrive at the boundaries of the flake. The zigzag edges of the flake give a different spectrum from that of the square flake. Generally the zigzag boundary graphene has edge states and provides edge channels for electrons. However, away from the Dirac point there exists antiresonant points, which makes the edge channels closed~\cite{jia}.  Our results confirm this by that there is no quasi-stationary current in this range. The impulse currents reach a maximum value and then decay to the stationary states. Compared to the square flake (the bottom panel of Fig.~\ref{fig:squaretransient}), there is no quasi-stationary plateau curves for the honeycomb flake (the top panel). In this case the honeycomb flake does not support a quasi-stationary current. Instead, the currents approach zero in a fluctuation fashion. The smaller the flake is, the larger the fluctuation is. The conclusion is that the different configurations of the flakes show different behaviors in the transient regime.

\section{The transient current through a square flake in the presence of disorder}

\begin{figure}
\begin{center}
\includegraphics[width=0.7\textwidth]{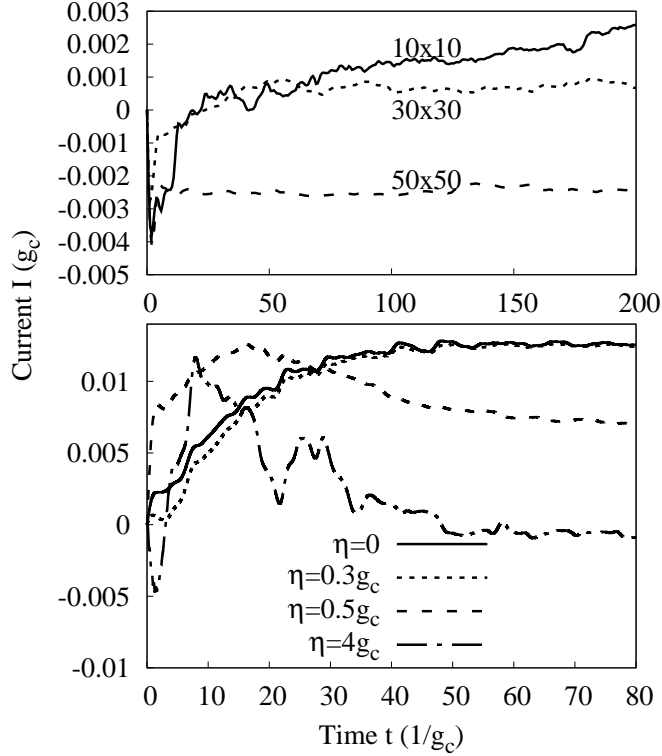}
\caption{[Bottom panel] The transient current through a square flake of size 5x5 in the presence of disorder. [Top panel] The transient current through square flakes of different sizes as $\eta=0.5g_c$. The voltage bias is set to $V=0.1g_c$.\label{fig:squaredisorder}}
\end{center}
\end{figure}
In the presence of disorder the transient currents through a square flake are shown in Fig.~\ref{fig:squaredisorder}, in which the bottom panel shows the currents at different disorder strength $\eta$ while the top panel shows the currents in the flakes of different sizes at fixed disorder strength $\eta=0.5g_c$. We find that the quasi-stationary currents disappear as the strength of disorder increases, which is caused by the random scattering and the increase of the resistance. The disappearance of the quasi-stationary currents does not occur abruptly. Instead, it gradually disappears as $\eta$ increase (the curve titled $\eta=0.3g_c$ is close to that titled $\eta=0$ in Fig.~\ref{fig:squaredisorder}).

As the strength of disorder increases, the relaxation time of current increases too. Without disorder, the current has almost relaxed to its stationary value as long as $80/g_c$. However, the current (see the curve titled $\eta=4g_c$ in Fig.~\ref{fig:squaredisorder}) relaxes much slower in the presence of strong disorder and does not reach the steady states even at $t=80/g_c$.

As expected from classical or quantum point of view, the relaxation time increases with the size of the flake since the relaxation time is related to the resistance and capacitance or inductance. The larger the flake is, the larger the relaxation time is needed. From the top panel of Fig.~\ref{fig:squaredisorder} we see that the transient current of the flake $50\times 50$ keeps negative value even at $t=200/g_c$, which obviously does not relax since the stationary current must be positive in these systems. In this case one needs to perform longer computational time to obtain the stationary current which is not necessary in this work.

\section{The stationary currents}

\begin{figure}
\begin{center}
\includegraphics[width=0.7\textwidth]{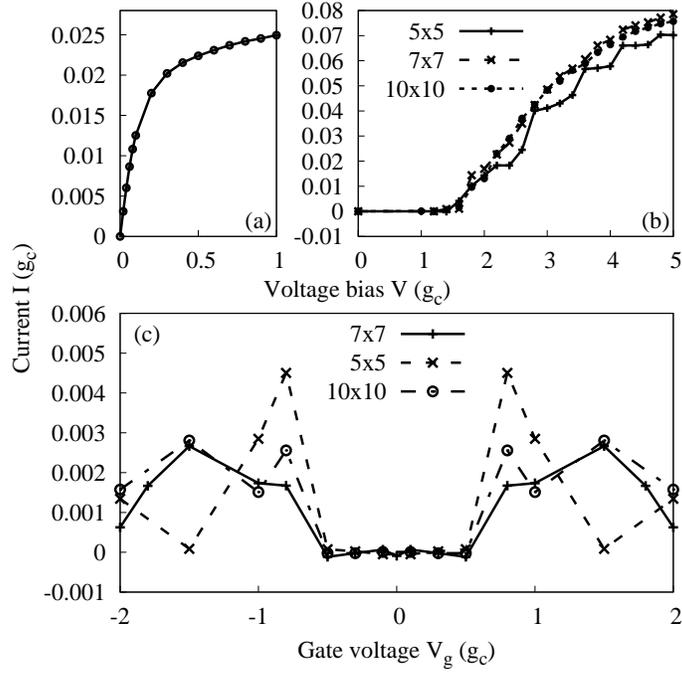}
\caption{(a) The stationary current through a square flake as a function of the voltage bias $V$. The flake size is $5\times5$. And the gate voltage is set to $V_g=0$.  (b) The currents through the honeycomb flakes of the size $ 5\times5, 7\times7$ and $10\times10$ as a function of the voltage bias. The gate voltage is set to zero. (c) The currents through the honeycomb flakes as a function of the gate voltage $V_g$ when the voltage bias is $V=0.1g_c$. \label{fig:squaresta}}
\end{center}
\end{figure}
We study the stationary currents by the numerical operator method for not too large flakes. In this section, we discuss the stationary currents through the square and honeycomb flakes in the absence of disorder.

First we discuss the current-voltage (I-V) curve of the square flake (see the panel (a) of Fig.~\ref{fig:squaresta}). The I-V curve of the square flake is found to be an inverse tangent function. The current increases smoothly from zero to the saturation value, and has the same feature as that of the single impurity Anderson model without interactions. This fact indicates that the density spectrum of a square flake is a Lorentzian function, which is the same as that of a single impurity site. This feature is expected since the density of states of a square flake is gapless at the Fermi energy, and at the same time, all its single-particle eigenmodes are coupled to the leads at the same strength.

However, the I-V curve of a honeycomb flake is much different. In the panel (b) of Fig.~\ref{fig:squaresta}, we clearly see a threshold voltage which is a little bigger than $g_c$. At small voltage bias the current is forbidden. This fact indicates that for the diamond shape of flakes with zigzag edges there exists an energy gap around the Dirac point. Exceeding the threshold voltage the current channels are open. In a similar way the current shows step characteristics which demonstrate the discrete quantum channels away from the Dirac point (see the curve titled $5\times5$ in the panel (b) of Fig.~\ref{fig:squaresta}). Our result also indicates that the spectral density of the honeycomb flake is not a Lorentzian function but more complicated by the behavior of Dirac fermions.

We verify the energy gap in the spectral density by studying the current in the linear response regime as a function of the gate voltage, which directly shows the shape of the spectral density of the flake. In the panel (c) of Fig.~\ref{fig:squaresta}, we show the current as a function of the gate voltage $V_g$. As $V$ is small the energy gap is clear.

\section{Conclusions}

In summary, we present the numerical operator method to efficiently deal with the real time dynamics of currents through nanostructures with or without disorder beyond the linear response regime. As examples, we use this method to study the transient and stationary currents through the square and honeycomb flakes. Both the transient and stationary currents are obtained and different properties are found for these two kinds  of flakes due to the difference of their topologies. In the square flake there exist quasi-stationary currents, while they do not exist in the honeycomb flake. In the square flake the stationary current increases monotonically with the voltage bias. While in the honeycomb flake we find a threshold voltage in the I-V curve, below which the current is forbidden, indicating an energy gap at the Dirac point. This gap is then verified by the current as a function of the gate voltage. In the presence of disorder, the quasi-stationary current is destroyed gradually as the disorder strength increasing, due to the random scattering and the increase of the relaxation time by the presence of disorder.

\section*{Acknowledgement}

We wish to acknowledge the support of NSFC (Grant. No. 10674116) and PCSIRT (Grant No. IRT0754), and the support of Department of Physics, Zhejiang University and ZJEDU Y201226145.








\end{document}